\documentclass[reprint,cha,superscriptaddress]{revtex4-1}

\usepackage{graphicx}
\usepackage{amssymb}
\usepackage{placeins}
\usepackage{fancyhdr}
\fancypagestyle{firststyle}
{
   \fancyhf{}
   \lhead{http://ieeexplore.ieee.org/document/7856194/}

}
\usepackage[T1]{fontenc}
\usepackage{color}
\usepackage{url}
\usepackage{rotating}
\usepackage[squaren]{SIunits}
\usepackage{amsmath}
\usepackage{wasysym}
\usepackage[latin1]{inputenc}
\usepackage[T1]{fontenc}
\usepackage{ae,aecompl}
\usepackage{xcolor}

\usepackage[font=footnotesize,format=plain,justification=justified,singlelinecheck=false]{caption}
\renewcommand{\sec}[1]{Section~\ref{sec:#1}}

\newcommand{\eqna}{\begin{eqnarray}}
\newcommand{\eqne}{\end{eqnarray}}
\newcommand{\eqnaa}{\begin{align*}}
\newcommand{\eqnee}{\end{align*}}

\definecolor{darkgreen}{rgb}{0,0.5,0}

\begin{document} 

\title{Can Distribution Grids Significantly Contribute to Transmission Grids' Voltage Management?}
\author{Sabine Auer}
 \email{auer@pik-potsdam.de}
 \affiliation{Potsdam Institute for Climate Impact Research, 14412 Potsdam, Germany}
 \affiliation{Department of Physics, Humboldt University Berlin, 12489 Berlin, Germany}
\author{Florian Steinke}%
\affiliation{Siemens Corporate Technology,  81739  Munich,}
\author{Wang Chunsen}
\affiliation{RWTH Aachen  University, 52062  Aachen,} 
\author{Andrei Szabo}
\affiliation{Siemens Corporate Technology,  81739  Munich,}
\author{Rudolf Sollacher}
\affiliation{Siemens Corporate Technology,  81739  Munich,}%


\begin{abstract}
Power generation in Germany is currently transitioning from a system based on large, central, thermal power plants to one that heavily relies on small, decentral, mostly renewable power generators.
This development poses the question how transmission grids' reactive power demand for voltage management, covered by central power plants today, can be supplied in the future.

In this work, we estimate the future technical potential of such an approach for the whole of Germany.
For a 100\% renewable electricity scenario we set the possible reactive power supply in comparison with the reactive power requirements that are needed to realize the simulated future transmission grid power flows.
Since an exact calculation of distribution grids' reactive power potential is difficult due to the unavailability of detailed grid models on such scale, we optimistically estimate the potential by assuming a scaled, averaged distribution grid model connected to each of the transmission grid nodes.

We find that for all except a few transmission grid nodes, the required reactive power can be fully supplied from the modelled distribution grids. 
This implies that -- even if our estimate is overly optimistic -- distributed reactive power provisioning will be a technical solution for many future reactive power challenges.
\end{abstract} 

\pacs{Valid PACS appear here}

\maketitle


%
\thispagestyle{firststyle}

\section{Introduction}
\begin{figure}[t]%
\hspace{-0.05\textwidth}
\begin{minipage}[c]{0.67\columnwidth}
    \includegraphics[width=0.85\columnwidth]{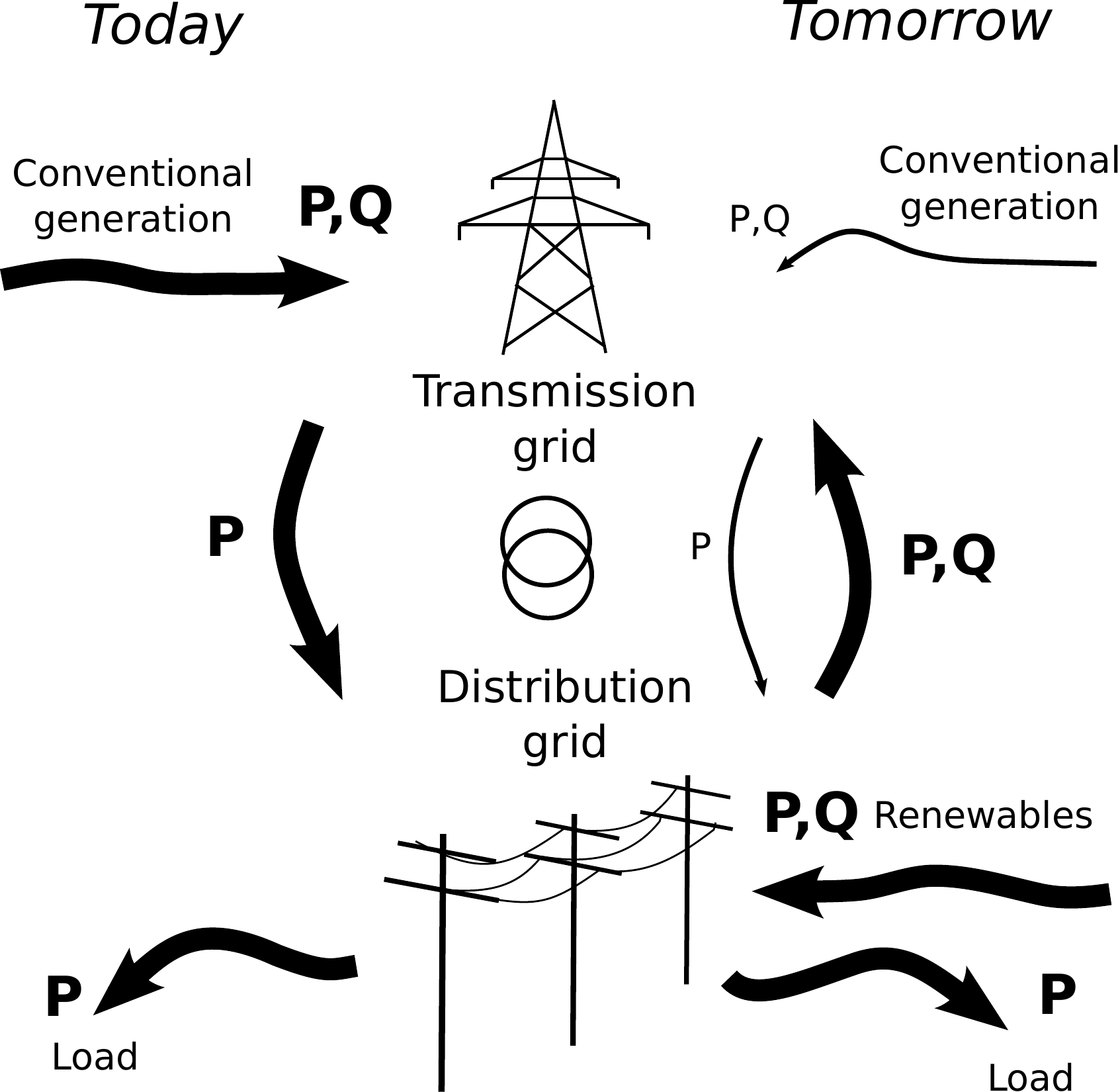}
  \end{minipage}
  \begin{minipage}[c]{0.32\columnwidth}
  \caption{Simplified scheme for the flow of active and reactive power for the scenarios ``today'' (conventional power generation at high grid levels) and ``tomorrow'' (renewable power generation in the distribution grid). P and Q are active and reactive power, respectively.}
     \label{fig:scheme_Qdemand}
  \end{minipage}
  \vspace{-0.02\textwidth}
\end{figure}

The increasing share of renewable energies (RE) in Germany shifts the paradigm of electricity generation from large, centralized power production to so-called distributed generation where producers are mainly installed in the distribution grid (see Fig. \ref{fig:scheme_Qdemand}). 
As the placement of RE generators like wind and solar power plants takes place according to their optimal yield instead of vicinity to power consumers, the distances of power transmission increases. 
Altogether, this results in a more stress for the power infrastructure, and thus the demand for reactive power increases at all grid levels. 
At the same time REs offer a large potential to generate reactive power with their well-controllable DC/AC converters.


Since reactive power is best re-compensated locally, these decentral generators are very plausible suppliers of reactive power in the DG. 
However, several studies \cite{dierkes2014impact,talavera2015vertical,sowa2015potential} have shown that reactive power can also be transported to DGs' connection with the transmission grid (TG) and have quantified its technical potential for exemplary DGs.
Several publications propose local automatic reactive power control approaches and tap changer coordination \cite{wang2015controlled,carvalho2008distributed,mamandur1981optimal,viawan2008voltage,turitsyn2011options,viawan2008voltage,demirok2011local} to stabilize voltage levels. In Kraiczy et. al.  \cite{kraiczy2015parameterization}, the authors underline the need for a grid code revision for an automated variable RE control which is in line with our research aim on quantifying the potential of such decentral reactive power provision. Talavera and Sowa et. al. \cite{talavera2015vertical, sowa2015potential} analyse the potential of reactive power exchange between MV- and HV networks with either fixed or arbitrary power factor and remotely controllable HV/MV tap changers. 

In this paper, we try to quantify the technical potential of reactive power from the DG on a country-wide scale, and in relation to possible reactive power demands. In our analysis we also include the HV grid, allow for arbitrary power factors at all voltage levels and  controllable LV/MV tap changers. 
To this aim, we use the future 100\% renewable scenario of the Kombikraftwerk 2 (KKW2) study \cite{kombikraftwerk} which entails a transmission grid model and values for the future power flows on each line for each hour of a reference simulation year.

A significant technical potential of the decentral (often RE) generators in covering TGs reactive power demands would be a strong argument in favor of investment in 
a) a smart-grid communications infrastructure between the grid operator and decentral generators, 
b) reactive power optimized DGs employing e.g. additional tap-changeable transformers,  
as well as c) the corresponding software tools to correctly control such potential from grid operators' control rooms.
In this work we will focus on the technical potential and leave an economic discussion to the future. 
The final decision process will include dual-use arguments for such infrastructure as well as the cost for the alternative, that is, additional, central reactive power compensation hardware.

In \sec{Model}, we describe the DG model, which we then apply to three prototypical test cases: a PV-dominated region, a demand- and a wind-oriented one, see \sec{experiments}, to validate our methodology.
In \sec{results_KKW2}, we then extend this analysis to the whole of Germany, quantifying for the future 100\% renewable scenario from KKW2 \cite{kombikraftwerk} how much of TG's overall reactive power demand can be satisfied from the DGs.
We discuss our findings in \ref{sec:discussion}.

\section{Model}\label{sec:Model}
We derive a scalable, average distribution grid model for Germany assuming full symmetry of tree topology. 
This is certainly not realistic, but it represents an optimistic estimate. 
We then use power flow computations to determine the minimal and maximal reactive power that can be provided at the connection point between transmission grid and distribution grid.
We apply these steps to each transmission grid node that is modeled in the 100\% renewable energy scenario for Germany 2050 from 
the Komikraftwerk 2 study \cite{kombikraftwerk}.
For this setting, we compare the possible reactive power generation with the demands resulting from the modeled transmission grid power flows from \cite{kombikraftwerk}.

\subsection{Scalable, Average German Distribution Grid Model}
For the purpose of estimating the potential of reactive power from the distribution grid on a Germany-wide scale, 
we assume that at every connection of the transmission grid, ultra-high voltage (UHV), to the distribution grid, high-voltage and below, a specific number of identical copies of an average German distribution grid is connected -- with fully symmetric topology. 
\begin{figure}[b]%
\centering
\includegraphics[width=0.95\columnwidth]{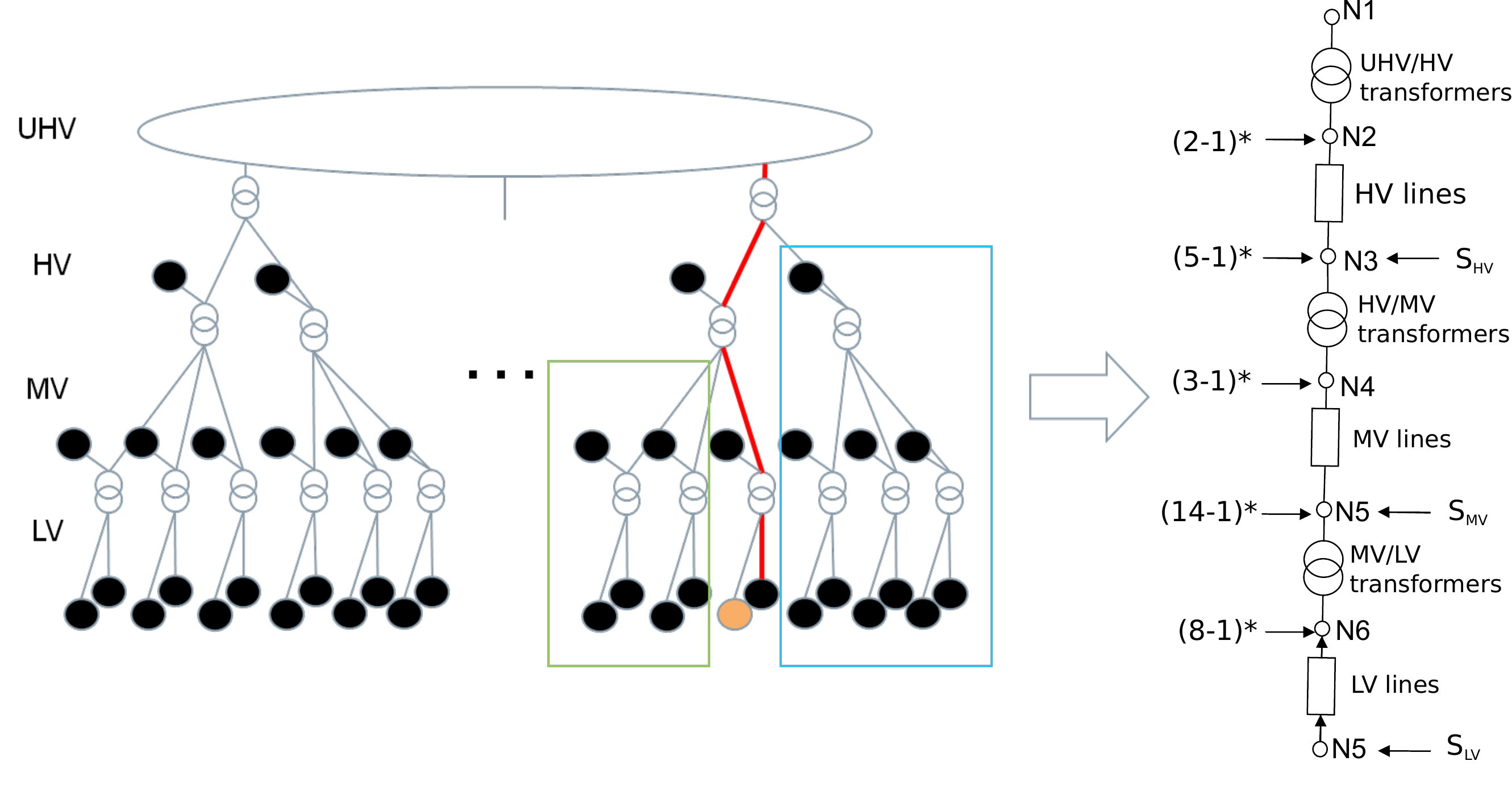}
\caption{Schematic illustration of grid topology reduction to a simplified chain-like structure due to full tree symmetry.}%
\label{fig:topologyReduction}%
\end{figure}
Since also the generation capacity as well as its actual values and the loads are assumed fully symmetric, the complex voltage state at each node on the same level of the grid will be identical in each load situation. We thus have to consider only one representative path from the UHV down to the LV level for our power flow calculations, see Fig. \ref{fig:topologyReduction} for a schematic illustration.  
 
Every grid level is represented by a link, realized via per unit (pu) calculations, with an impedance followed by a transformer with tap changer in the range +/-0.05 p.u. to connect to the neighboring grid level (see Fig. \ref{fig:topologyReduction}). 
The apparent power $S_i$ can be understood as the power infeed from one of the $N_i$ nodes at each node level in our simplified power grid. The injections at transformer level comes from the symmetric generation accumulation.         
Consumer loads are placed at the end of the lines. 
As in the MV and LV grid loads are rather distributed homogeneously along the power lines we take only half of the lines' typical length in order to get the same voltage drops along the lines \cite{dierkes2014impact}.

The topology of the average tree graph structure is given by the number of grid nodes which are connected by lines and transformers (see \ref{sec:app_gridparameters}).
The approximate proportionality ratios, corresponding to the tree graphs branching, then read 
\begin{equation}
\begin{array}{ccccccc}
1 &:2&: 3&: 14 &:9&: 8,  \end{array}
\label{eq:ratio_wo}
\end{equation} 
that is, for each UHV/HV transformer there exist 2 HV lines, 6 HV/MV transformers etc. 
Since Germany's peak load is about $80$ GW and there are $606$ UHV grid nodes \cite{SciGRIDv0.1}, each average distribution grid has a peak load of $132$ MW.

The real UHV nodes, however, have different peak consumer loads connected to them. Some nodes support large loads in big cities, other rather supply load-poor rural areas.
We thus rescale each node's generation capacities and loads to match one copy of our average grid model, perform the distribution grid power flow calculations, and then rescale its results to the original node size. Specifically, we introduce the norm factor $n_{load,i}$ for each UHV node $i$ as
\begin{equation}
n_{load,i} = \max_t(D_{i,t})/132MW, \forall i
\end{equation}
where $D_{i,t}$ is the consumer load at node $i$ at time $t$.
We then normalize the installed distribution power generation capacities $C_{i}$  at node $i$ with current actual power generation $P_{i,t}$ at time $t$ to our average grid model as
\begin{eqnarray}
D^{'}_{i,t}&&= D_{i,t}/n_{load,i},\\
C^{'}_{i}&&= C_{i}/n_{load,i},\\
P^{'}_{i,t}&&= P_{i,t}/n_{load,i}.
\end{eqnarray}
Moreover, if the distributed power generation capacity $C^{'}_{i}$ exceeds the $1.5$fold of the peak load of $132MW$ of the modeled distribution grid structure, it is unlikely that such generation capacity could be integrated into the existing grid without massive grid extensions.
We thus only considered the plausibly integrateable part below such thresholds and normalize the generation capacities a second time with factor $n_{cap,i}$ defined as 
\begin{equation}
n_{cap,i} = \begin{cases} C^{'}/(1.5\cdot 132)& C^{'}_{i}> 1.5\cdot 132 \\ 1 & C^{'}_{i}\leq 1.5\cdot 132 \end{cases},  
\end{equation}
yielding 
\begin{eqnarray}
C^{''}_{i}&&= C^{'}_{i}/n_{cap,i},\\
P^{''}_{i,t}&&= P^{'}_{i}/n_{cap,i}. 
\end{eqnarray}
After the power flow simulation described in the following sections, the computed reactive power exchanges at the UHV level $Q_{UHV,i}^{''},$ for $ C^{''}_i$ and $P^{''}_{i,t}$ are scaled back by 
\begin{equation}
Q_{i,t}=Q_{i,t}^{''}\cdot n_{cap,i}\cdot n_{load,i}.
\end{equation}

\subsection{Reactive Power Optimization}
To maximize/minimize the reactive power provision at the connection between HV and UHV, we use a standard interior-point method for non-linear optimization.
The independent variables are the reactive power infeeds at the different voltage levels, namely  $Q_{HV},Q_{MV}$ and $Q_{LV}$.
The objective function is 
\begin{equation}
c(Q_{HV},Q_{MV},Q_{LV}) = \pm Q_{UHV}.
\end{equation} 
The sign of the objective, positive or negative, determines whether capacitive or inductive reactive power is maximized, respectively.
$Q_{UHV}$ is calculated using the well-known forward/backward sweep method for solving the power flow equations in tree grids \cite{eminoglu2008distribution}, in each iteration of the optimization. 
Constraints to the optimization problem are
\begin{itemize}
	\item apparent power of the generators
	 $Q^2<C^2-P^2$ and
	\item maximally allowed voltage fluctuations 
	$\Delta U =\pm 10\%$ at each grid node.
\end{itemize} 
Tap changer position of all transformers were assumed to be adjustable to the discrete values [0.95 1.0 1.05], and we chose the optimal tap settings from any possible combination of these options.  
We assume a fixed power factor of 0.95 for loads.

In this study, line capacity constraints were neglected as previous studies have indicated that too high or too low voltages are typically the first problem seen in distribution grids with massive decentral generation \cite{lodl2011operation}. We did, however, check the transformer load limits after the power flow computation and found them to be less crucial than the voltage limits, again similar to \cite{lodl2011operation}.


\subsection{Germany-wide Analysis}
In the Kombikraftwerk 2 study \cite{kombikraftwerk} a 2050 scenario of the German power system was developed with 100\% renewable energy sources. 
The study details the installed capacities of renewables, storages and backup power plants to the level of UHV transmission grid nodes in Germany, and developes a dispatch for the full system for each hour of one meteorological weather year.
The resulting power flows in the modeled transmission grid, a slightly extended version of the grid proposed in the Netzentwicklungsplan 2012 \cite{Zustand13}, and a Q(U) droop control law lead to reactive power demands at each node for each point in time.

Note that the reactive power demand of the modeled TG is mostly capacitive. This is because the modeled grid is highly expanded in comparison to today, but the additional lines are only rarely used and thus thus behave in a capacitive way.

We use the reactive power demands from \cite{kombikraftwerk} and match them with our estimate of the possible Q generation from underlying distribution grids.
Therefore, we also used their distributed generation capacities for each TG node and the split to the different voltage levels mentioned therein.
To reduce computational effort we condensed the $8760$ given time steps with kmeans from Matlab \cite{MATLAB:2014} into $30$ time clusters that cover a wide range of power flow situations.
To account for the local exchange of reactive power between neighboring TG nodes and to reduce the impact of local grid modeling errors (e.g. falsely assigned transformators between the 220kV and 380kV TG levels), we aggregated the possible Q generations over a shortest path radius of $\leq 30$ km.
TG nodes outside Germany and nodes without any consumption load were excluded from our analysis.

\section{Results for Prototypical Distribution Grids}\label{sec:experiments}

\begin{figure}[t!]
\centering
(a) Passau\\
		\includegraphics[width=0.443\textwidth]{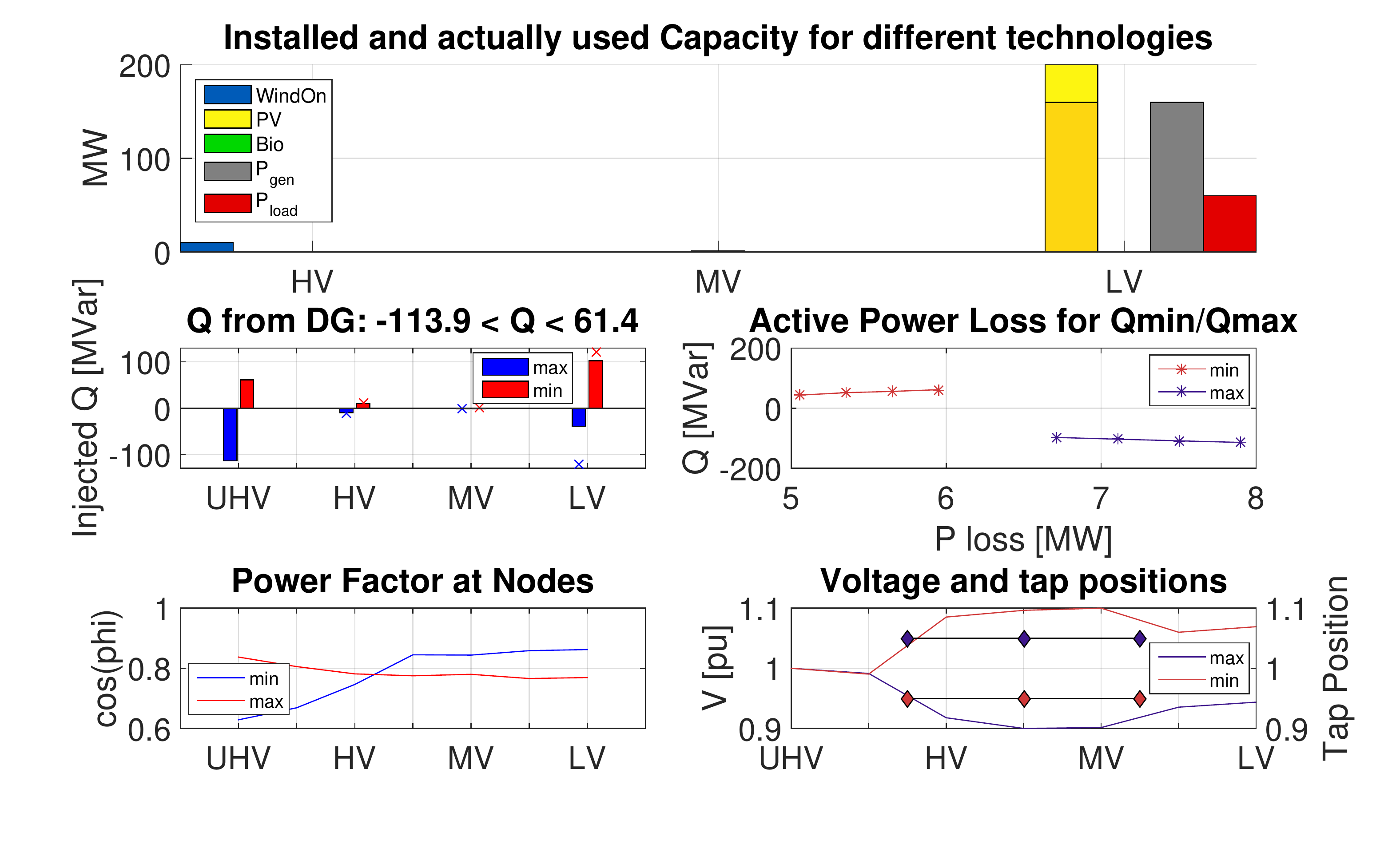}
		\vspace{-0.02\textwidth}\\
(b) Munich\\
		\includegraphics[width=0.443\textwidth]{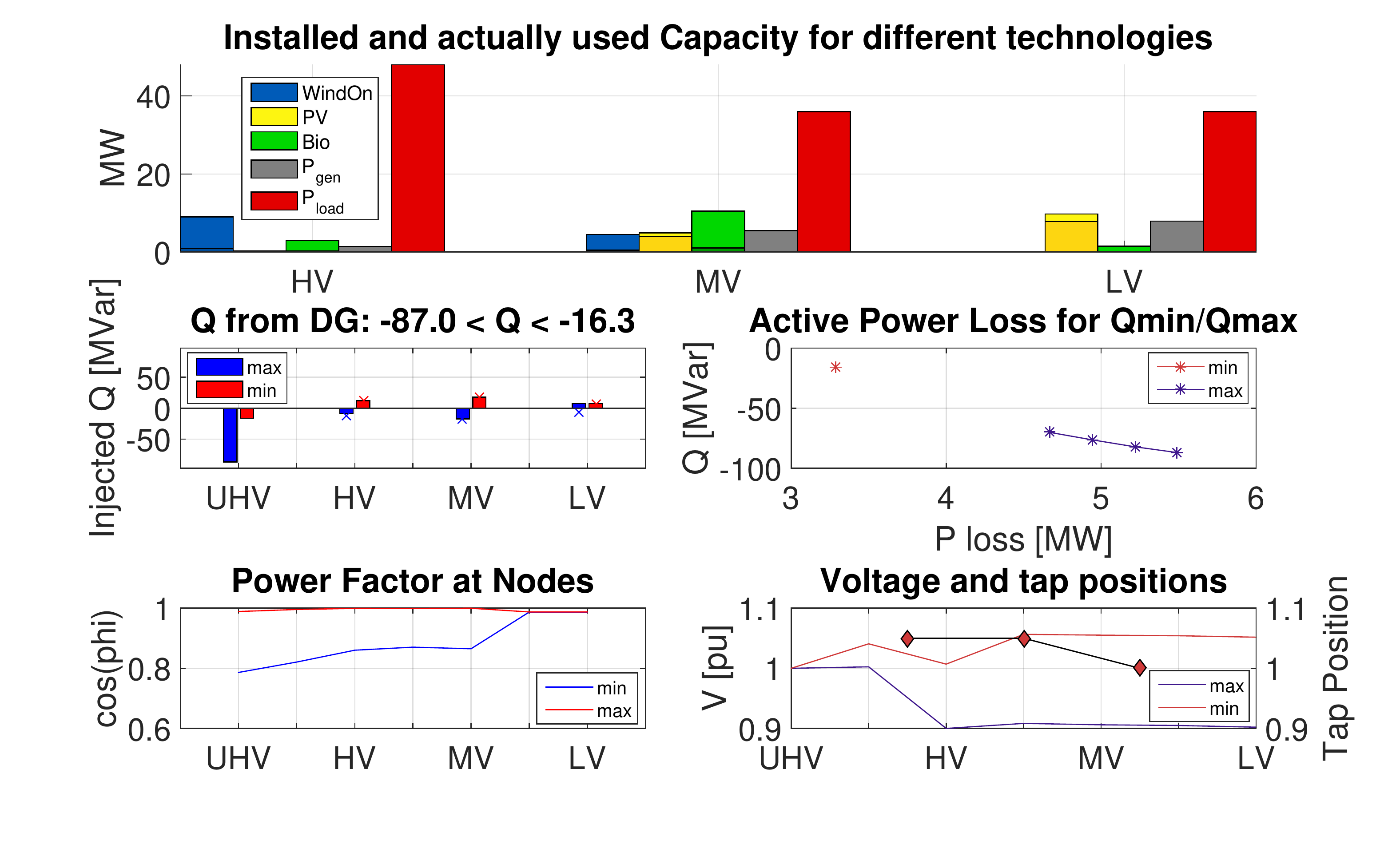}\vspace{-0.02\textwidth}\\
(c) Goerlitz \\
		\includegraphics[width=0.443\textwidth]{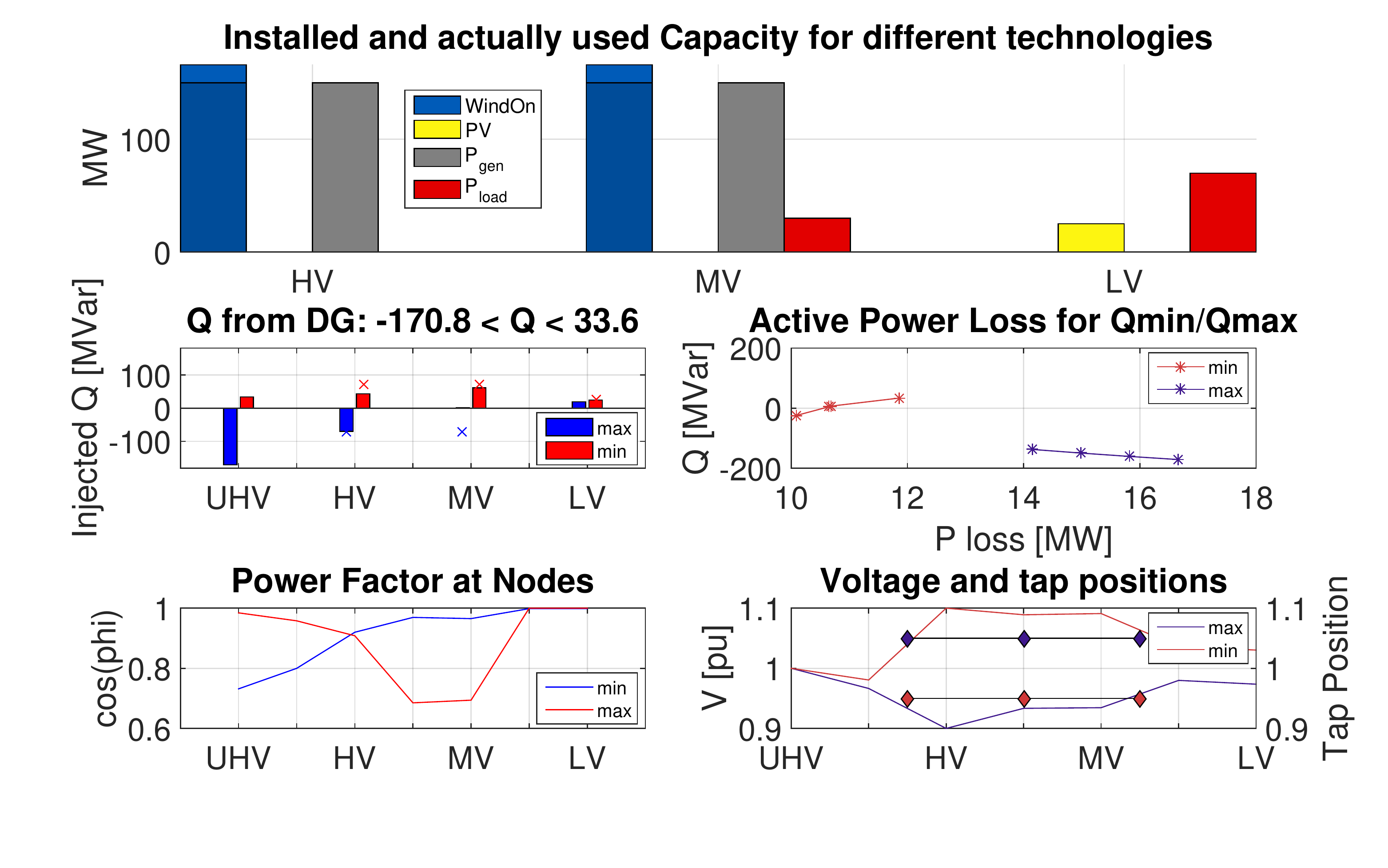}\vspace{-0.02\textwidth}\\
		\small{\caption{For each of the three prototypical grids we show the following plots:
(Upper panel) generation capacities and actual generation (light and dark colors respectively) for different technologies and grid levels. Additionally, overall generation and consumer load.
(Mid-left panel) The bars show the total reactive power generation of the generators per grid level if the reactive power transfer to the TG is maximized (blue) or minimized (red).  
The UHV values show the generation as seen from a perspective of the TG (inverse sign relative to the DG's view that is used for the other grid levels).
Crosses mark the reactive power potential for each grid level as derived from the installed capacities and their current usage.
Negative $Q$ corresponds to inductive and positive $Q$ to capacitive reactive power generation. 
(Mid-right panel) The maximally/minimally possible Q transfer to the TG is shown for different constraints on the implied total active power losses in the distribution grid.
(Lower left panel) The power factor of the injected power (generation minus consumption) is shown. 
(Lower right panel) The right axis encodes the optimal transformer tap positions that minimize and maximize reactive power transfer to the TG. The left axis denotes the voltage magnitude in the different grid levels relative to nominal voltage for Q maximization/minimization.}}
	\label{fig:allscen}
\end{figure}

We construct three distribution grids with prototypical mixes of generation capacities and consumption. 
We investigate their potential for reactive power support from the DG to the TG and the limitations thereof, for demanding load/generation situations. 
The prototypes are code-named and characterized as follows:
\begin{enumerate}
	\item[-]\emph{Passau}: sunny midday in a rural area with small load and large PV generation in the LV grid 
	\item[-]\emph{Munich}: evening in an urban area with large load and little decentral generation
	\item[-]\emph{Goerlitz}: windy afternoon in a rural area with small load and large wind generation in the MV and HV grid
\end{enumerate}
The results of our simulations are shown in Fig. \ref{fig:allscen}.

\subsection{Passau}

The constructed prototype \emph{Passau} shown in Fig. \ref{fig:allscen}(a) has an active power flow of about 100 MW from the DG to the TG, 
resulting from 160 MW solar generation and 60 MW load (relative to an assumed peak load of 132 MW for the grid).
With 80\% active power usage of the installed PV capacities there is still a large reactive power potential available. 
Hence, it is possible to generate both capacitive and inductive reactive power at the connection from the DG to the TG.
The full Q potential of the generators in the DG can, however, not be exploited.
The limiting factors are the MV voltage limits -- under the assumption that transformers at all grid levels are switchable. 
The transformer tap positions are set to extreme values.
Despite the restricted usage of the inductive reactive power potential of the PV plants inthe LV, the possible inductive Q delivery of the DG to the TG grid is large. 
This is due tot the inductive Q generation of loads and the grid itself, shifting the overall Q exchange at UHV level to negative values.
Depending on the desired Q transfer from the DG to the TG grid, DG's active power losses vary by up to 3MW, which gives a loss to Q generation ratio of 1 to 20.

\subsection{Munich}

The \emph{Munich} prototype represents the classical power flow situation, with active power transferred from the TG / UHV level to the HV, MV and LV connected loads. 
As shown in Fig. \ref{fig:allscen}(b), few decentral generators lead to a rather small potential for local reactive power generation. 
Hence, the total Q potential of the DG as seen from the TG is always inductive, as is typical for load regions today. 
Aside from the LV inductive Q generation pushing voltage down to its lower limits, the existing Q potential can be fully used. 
For this prototype, over voltages are never an issue.
The main task of the assumed tap changers is that of loss reduction, since high voltages at constant power transmission reduce the electric currents in the power grid and correspondingly the losses.

\subsection{Goerlitz}

The prototype named \emph{Goerlitz} has large power flows from the HV to the UHV level as well as to the LV level. 
Satisfying 100 MW of electricity demand in the LV grid, the assumed local wind generation still exports an effective active power of 200 MW to the transmission grid. 
Fig. \ref{fig:allscen}(c) shows how this prototype DG is mainly able to generate inductive reactive power as seen from the TG perspective. 
In this model case with lots of HV generators, Q generation is constrained by the HV voltage limits first.
Moreover, the existing potential for inductive Q generation cannot be fully used, especially in the MV level, 
since loss-minimization implies to exploit reactive power capabilities close to the TG first. 
In order to stay within the voltage limits, the LV Q infeed counteracts with the HV one, e.g. to prevent the HV voltage to drop below its limits for an inductive Q exchange at UHV level, it is necessary to feed in capacitive reactive power at LV level.
Tap changers are set to their extreme positions. 
Losses are high since great amounts of power (larger than the 132MW peak load that grid is designed for) flow from the DG to the TG.

\section{Germany-Wide Analysis}\label{sec:results_KKW2}

\begin{figure}[t!]
\centering
\includegraphics[width=0.23\textwidth]{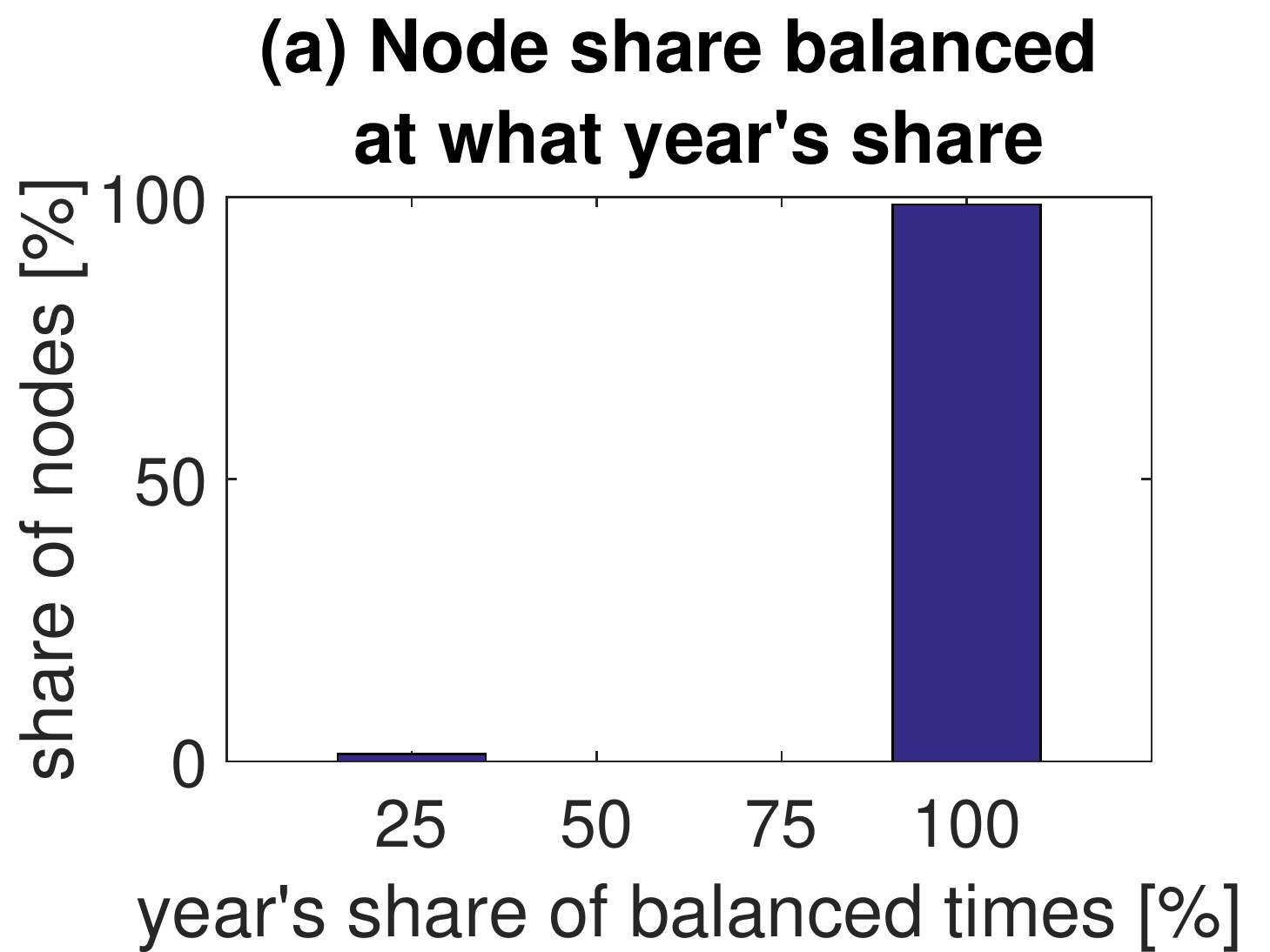}
\includegraphics[width=0.23\textwidth]{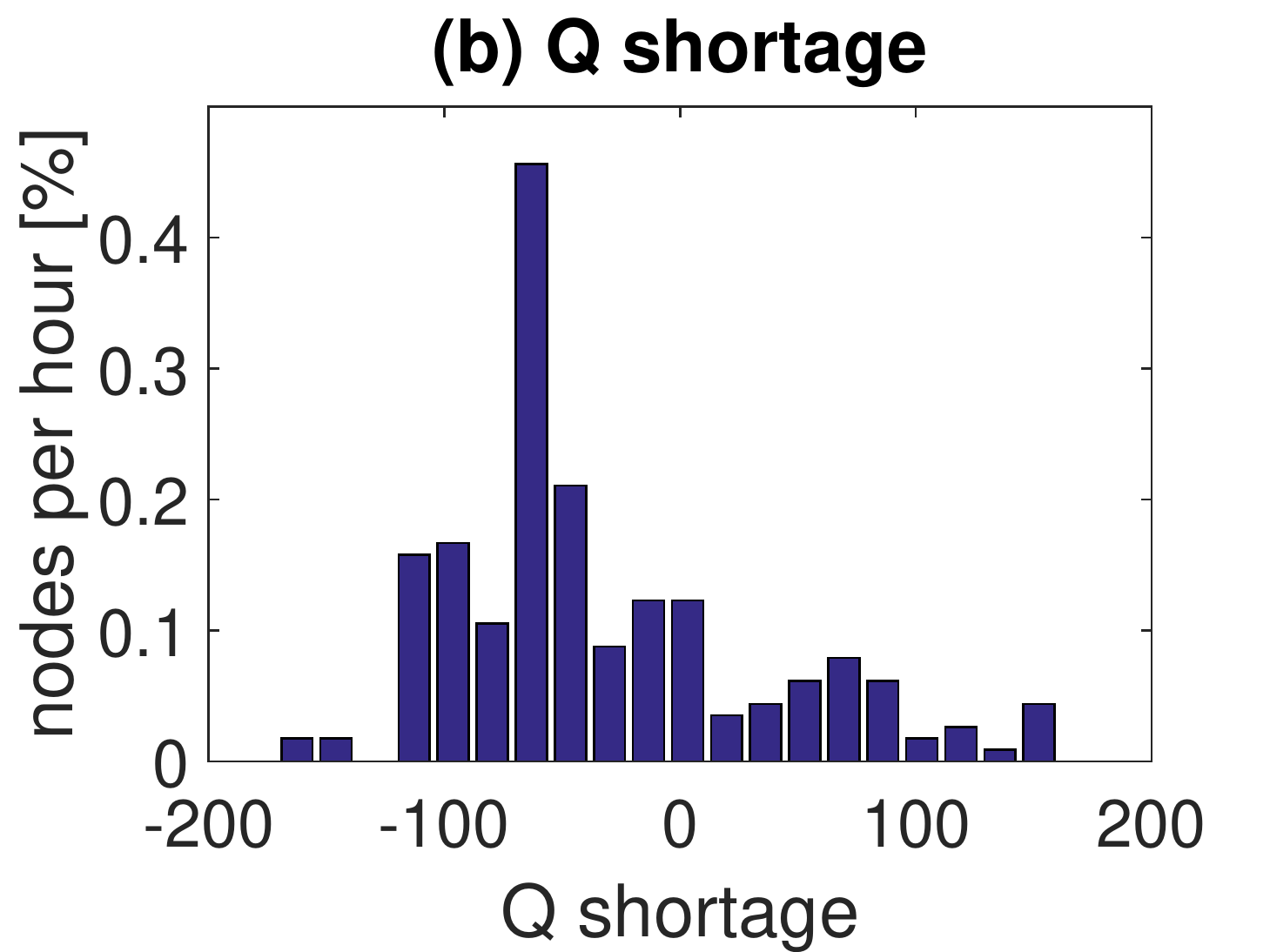}	\\
\vspace{0.01\textwidth}
\includegraphics[width=0.23\textwidth]{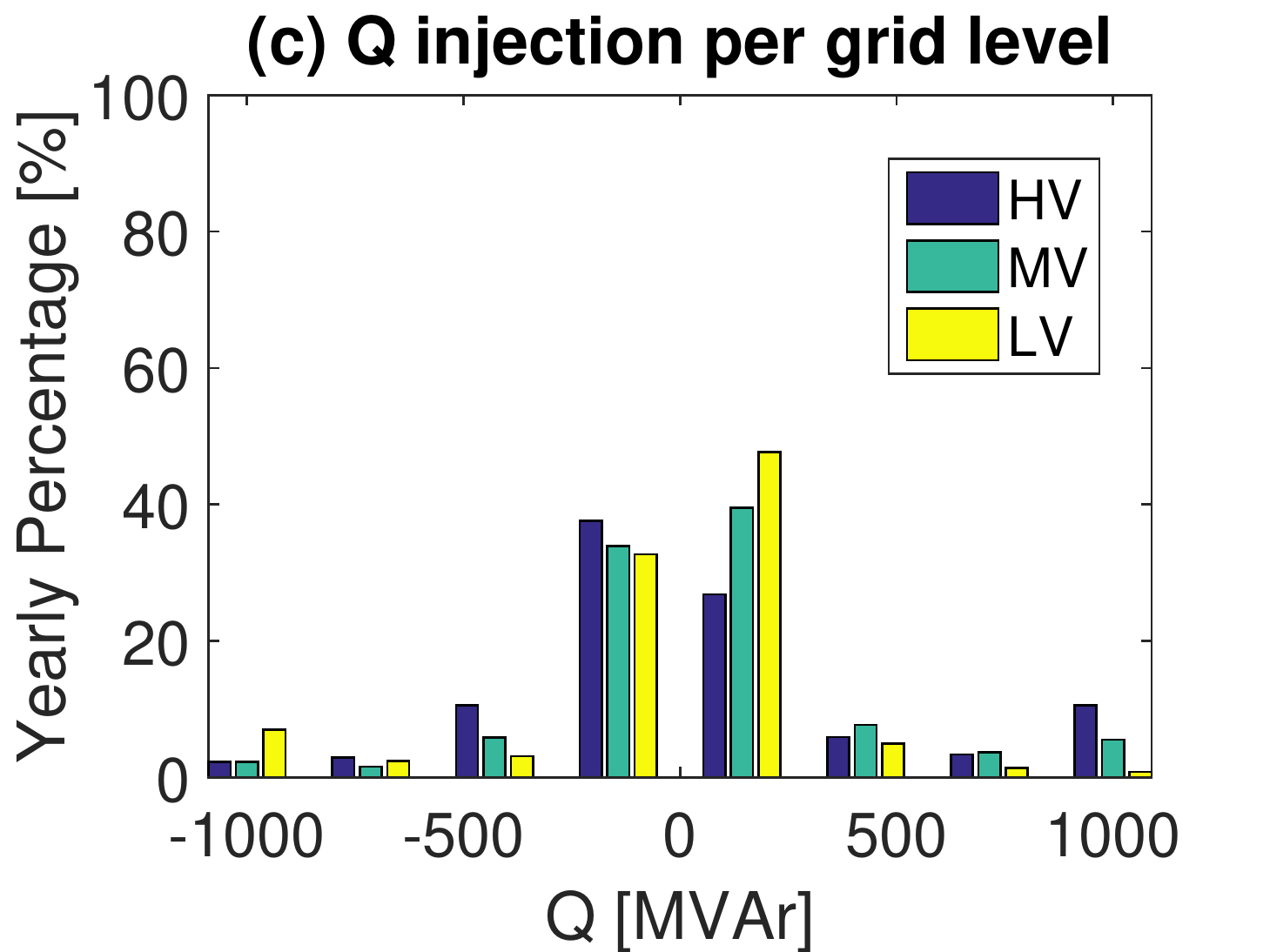}
\includegraphics[width=0.23 \textwidth]{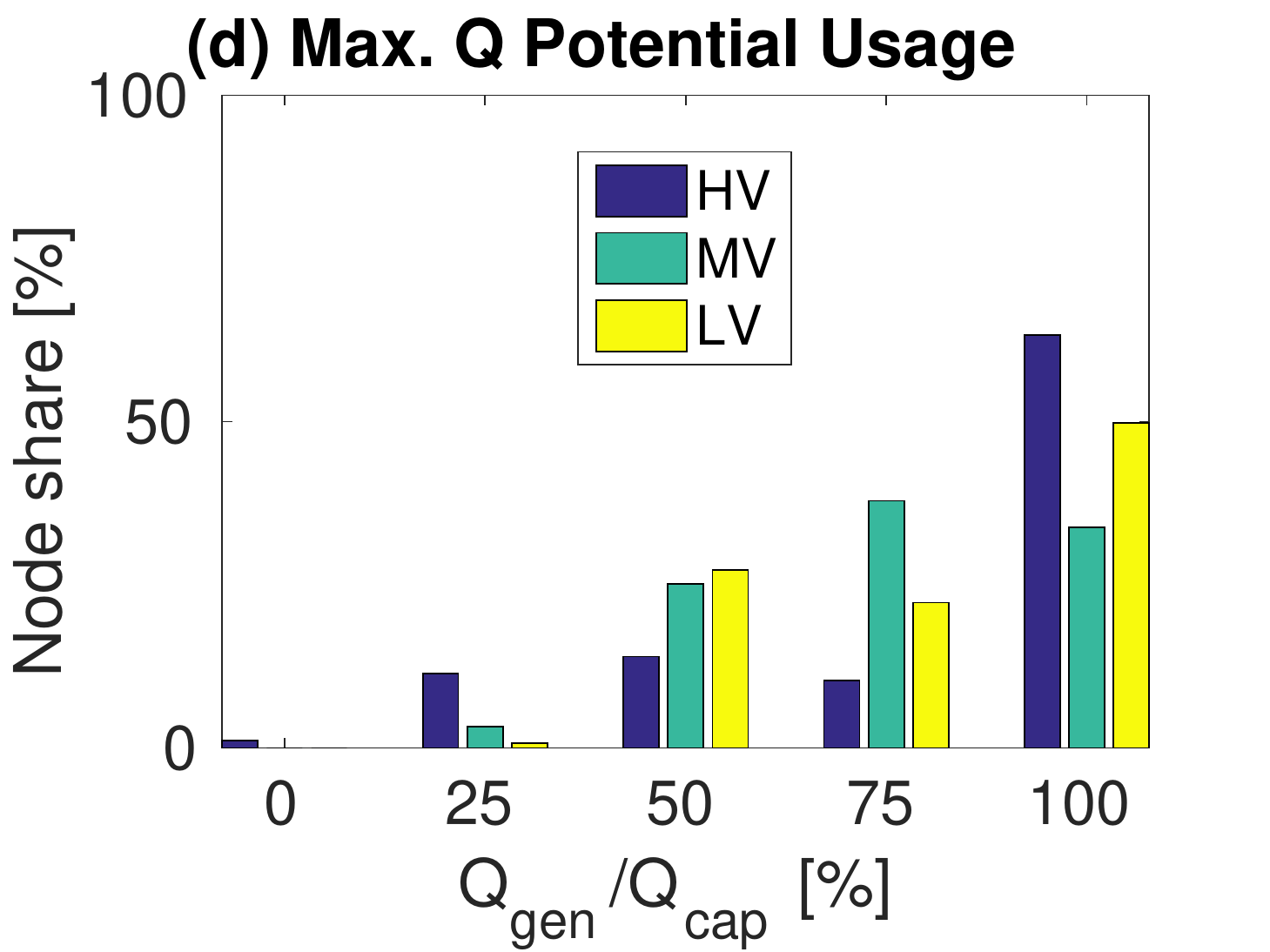}
\caption{(a) Fraction of nodes which can locally be Q-balanced at least a given time-percentage of the year, i.e. the bar at 100\% shows the fraction nodes that can always be balanced (98\% of the nodes). (b) Histogram plot of deviation from Q demand of the transmission grid for what share of nodes per time step. The nodes' Q generation at UHV level was aggregated within a range of shortest path smaller than $30$ km.
(c) Histogram of reactive power injection from different grid levels together with (d) showing how many nodes used what share of their available Q potential per grid level. E.g. 3\% of all nodes used their LV Q potential on average 100\%.}
	\label{fig:KKW_barplot}
\end{figure}

Setting the local reactive power potential from the DGs in relation to the reactive power demands of the TG for the Germany-wide 100\% RE scenario from the KKW2 study \cite{kombikraftwerk} yields the results shown in Fig. \ref{fig:KKW_barplot}.
For almost all TG nodes (98\%) the local reactive power demand can be compensated with reactive power generation from the distribution grids in the near neighborhood, i.e. from within 30km, see Fig. \ref{fig:KKW_barplot}(a).

\begin{figure}
\centering	
\includegraphics[width=0.5\textwidth]{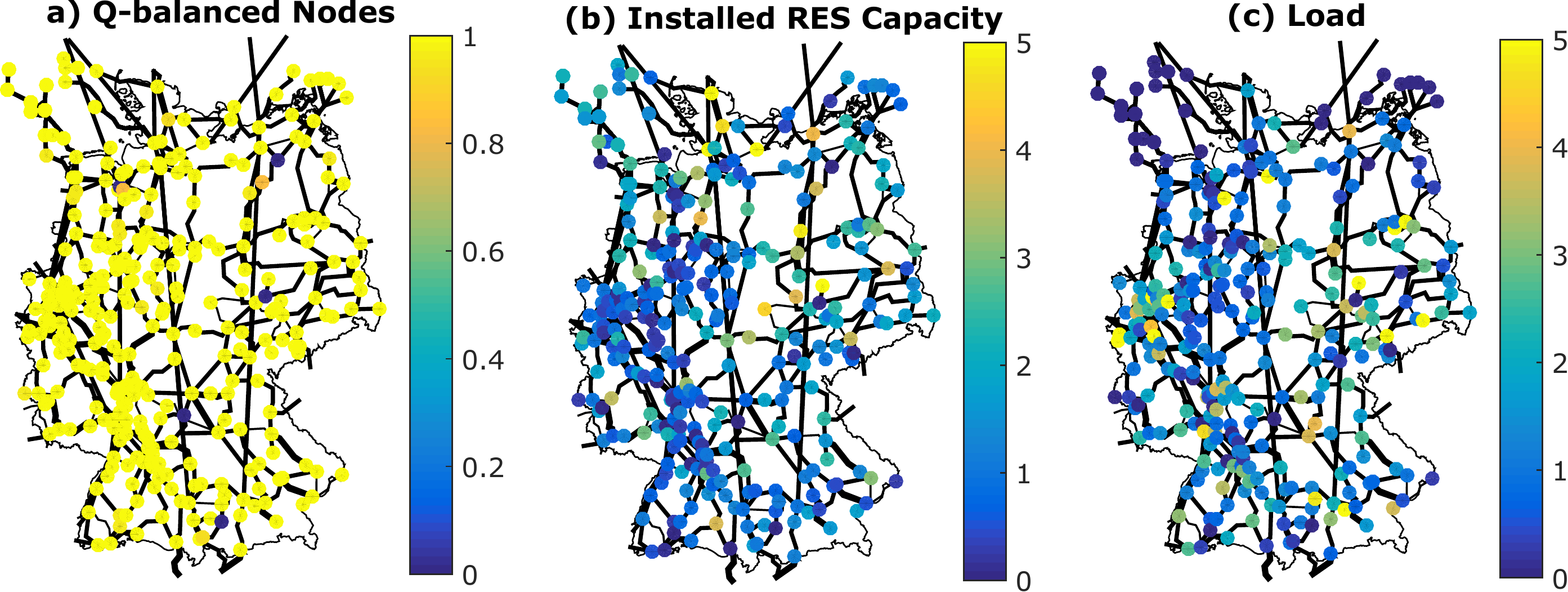}
\caption{
Results of the Germany-wide demand/supply potential analysis based on the Kombikraftwerk 2 scenario \cite{kombikraftwerk}:
(a) Colors encode for each TG node the share of time steps where its local reactive power demand can be balanced with local DGs' reactive power generation potential.
(b), (c)  Heat map of locally installed RES capacity and average consumption load relative to their Germany-wide average values. Values are capped at 5.
(d) Fraction of nodes which can locally be Q-balanced at least a given time-percentage of the year, i.e. the bar at 100\% shows the fraction nodes that can always be balanced (98\% of the nodes).}
	\label{fig:KKW_map}
\end{figure}

For the remaining 2\% of TG nodes for which the local reactive power demand cannot be balanced with reactive power generation from nearby DGs at most 200MVAr of additional capacitive or inductive reactive power are needed for local compensation, see Fig. \ref{fig:KKW_barplot}(b). 
The Q shortage is strongly skewed towards negative values and thus, there is a larger lack in inductive than capacitive reactive power.
The unbalanced nodes are geographically isolated (see Fig. \ref{fig:KKW_map}). 
They also coincide with nodes that have little or no decentral generation capacities and thus very limited local flexibility.

To examine what grid level contributes how much to the reactive power balance, the histograms in Fig. \ref{fig:KKW_barplot}(c) and (d) show how much Q is injected from each grid level and which share of Q capacity this maximally corresponds to over the year. 
According to Fig. \ref{fig:KKW_barplot}(c) all voltage levels often contribute to some capacitive and inductive reactive power production. 
Large quantities of inductive reactive power are occasionally provided by LV levels 
whereas MV and especially HV grid levels succeed at supplying large quantities of capacitive power.  
The maximal utilization rate of the existing Q capacity, see Fig. \ref{fig:KKW_barplot}(d), is about 50\% for all grid level.

\begin{figure}[h!]
\begin{minipage}[c]{0.65\columnwidth}
\includegraphics[width=0.9\columnwidth]{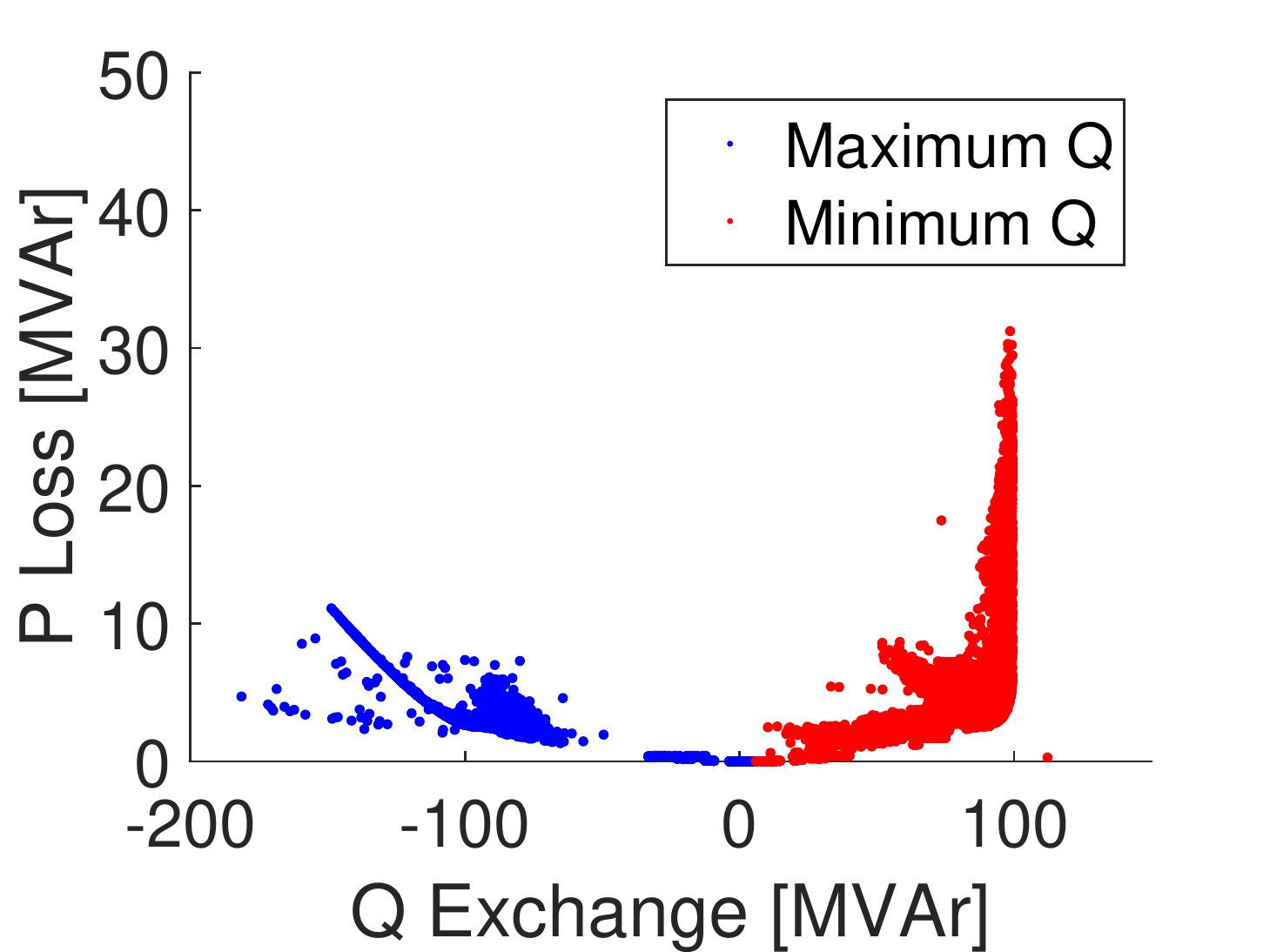}	
\end{minipage}\hspace{-0.02\textwidth}
 \begin{minipage}[c]{0.35\columnwidth}
\caption{Plot of active power losses over reactive power and for average UHV nodes with $132$ MW peak load. One data point corresponds to one simulated time cluster of one single node.}
	\label{fig:losses}
\end{minipage}
\vspace{-0.025\textwidth}
\end{figure}

In the previous section, we have seen that decentral Q-generation comes at the cost of increased active power in the DGs. 
The plot of active power loss over optimized reactive power for different average UHV nodes (for normalized grids with scaled loads and power production capacities), see Fig.\ref{fig:losses}, shows that active power losses are considerably higher for capacitive Q generation than for inductive one.
There is a sharp boundary for capacitive Q generation at $Q=100$ MVAr (for an average DG designed for 132 MW peak load), close to which the losses shoot up. 
The loss to Q generation ratios then may rise rises to the order of 1:3.
On the other hand, if one limits oneself to the range $-100<Q<70$ MVA $P_{loss}$, the additional losses from decentral reactive power generation in the DGs do not seem a pressing issue.

\section{Conclusion and Discussion}\label{sec:discussion}

Our study shows for a 100\% RE scenario for Germany that decentral reactive power from the DG is able to satisfy almost all reactive power demand from the TG. 
This means that DGs enabled to achieve this technical potential could be \emph{the} major tool for TG operators to manage the voltage in the TGs.

Reactive power demands that cannot be met by underlying DGs are few in our study, and they are locally isolated.
Consequently, few additional reactive power compensation devices in the TGs should be able to alleviate the remaining issues. These could be remaining large conventional power plants, HVDC converters, or special purpose compensation.\bigskip

To enable the proposed exploitation of Q compensation potential in the existing hardware of decentral RE generators,
it would be necessary to activate about half of the decentral power plant potential.
This means that their inverters would have to be able to work with arbitrary power factors and their Q generation would have to controllable via online communication (It needs to be checked whether fixed Q value or fixed droops would also do the job. In our study we have assumed centrally optimized setting for each situation).
Moreover, the DG voltage levels would have to be tightly controlled and thus observability in the DGs would have to be increase significantly.
Lastly, we have assumed tap changeable transformators at all grid levels. 
Many of these things, however, are already underway for different reasons, e.g. avoiding over-voltages in the DG due to RE infeed -- independent of additional Q generation.
A dual use for providing reactive power to the TG would thus be very welcome.

An economic counterargument may come from the increased DG losses that result from decentral Q generation. For extreme Q generation values these may be large. 
On the other hand, additional hardware for special purpose reactive power compensation in the DG grid is saved.
If macro-economically determined attractive the option to provide reactive power from the distribution grid requires new regulation to organize the financial compensation between the TG and the DG operator. First steps in this direction are explored in Switzerland \cite{Sch16}.\bigskip

Our technical results on DGs' Q potential may be optimistic for several reasons. 
We use a symmetric DG grid model, neglecting voltage  and corresponding Q limitations due to unsymmetrical branch loadings.
Our simulations are undertaken for time clusters with load and generation values representing the averages over the clustered hours. Hence, extreme load and generation patterns are not covered.
It is also not clear what structure future DG grid extensions that are required to integrate all the modeled REs will take. We have here assumed them to take the same structure as the existing ones.
However, even if all these arguments would in reality reduce the potential of reactive power from the DG by some fraction
the lever would still be a large one.

Moreover, while our DG model simplifications (symmetry, full power factor capability, remotely controllable Q set points and tap changers, full grid observability) are stronger than the one made in earlier work for exemplary networks, 
our numeric results are comparable.  
For the wind dominated distribution grid (eight rural 20-kV-networks) in \cite{talavera2015vertical}
the ranges of Q relative to peak load are -0.83 .. 0.83, for the same relative active power exchange as our \emph{Passau} scenario.
Our results of -1.29 .. 0.26 have a similar span, but are shifted towards inductive Q potential.
This may be due to the inclusion of load power factors in our calculations whereas in \cite{talavera2015vertical} this seems to be excluded from the reactive power exchange.
The taken DG model simplifications give us the possibility to examine a full country and to thereby compare the computed DG Q potential a plausible future TG Q demand in a detailed, consistent future energy scenario for Germany.

In future work, this study should be continued with more realistic DG models, e.g. as done with the random DG grid generator from the project \cite{ESDP}. Moreover, the economic evaluation also in comparison to the alternatives such as central Q compensation should be extended.

\section{Acknowledgements}
S.A. acknowledges gratefully the support of BMBF, CoNDyNet,  FK.  03SF0472A.

\bibliography{Literature}
\bibliographystyle{IEEEtran}
\appendix
\section{Appendices}

\subsection{Grid Parameters}\label{sec:app_gridparameters}
 
For the power lines the standard overhead and cable line values for the different voltage levels are taken, see Tab.~\ref{tab:links}. UHV/HV transformers have an overall capacity of $148$ GVA and are 606 in number \cite{SciGRIDv0.1}. This results in an average capacity of $244$ MVA. 
With 76,000 km HV lines \cite{denaverteilnetz} (ca. 94\% overhead lines) and a typical HV overhead line transmission capacity of 130 MVA on average there are $244/130\approx 2$ overhead lines per UHV/HV transformer. Then, average line length is $l_{HV}=76,000/(606\cdot 2))=62.7$ km.
Similarly, we proceed with the lower grid levels with $4080$ HV/MV transformers \cite{denaverteilnetz}, $523,468$km of MV lines (ca. 64\% overhead lines)\cite{Zustand13}.
Average line lengths are thus $l_{MV}=9.43$ km ($\tilde{l}_{MV}=0.5\cdot 9.43=4.72$ km as consumer loads are connected to the middle of the line).
Also, there are $460,321$ MV/LV transformers \cite{denaverteilnetz} and LV lines add up to $1,067,100$ km \cite{Zustand13}.

\begin{table}[h]
	\centering
	\begin{scriptsize}
		\begin{tabular*}{0.35\textwidth}{l|llllll}
			Transformers		& $UHV/HV$ & $HV/MV$  & $MV/LV$ \\ \hline 
			$C_j$ [MVA]			& 244  & 26	& 1  \\
			$U_j$ [kV]	&110	&	30		& 0.4	\\
			$U_k$ [pu]		  	& 0.14	&	0.1	& 0.14	&	\\
			$U_r$ [pu]			 & 0.005		&	0.005	& 0.005	&	\\
			$R$ [$\Omega/km$]		&0.2459		&	0.1731		& 8e-4\\
			$X$	[$\Omega/km$]	&6.8818	&	4.8431		& 2.24e-2\\
			$Y_{cr}$ [$km/\Omega$]	&0		&	0		& 0 \\
		\end{tabular*}
	\end{scriptsize}
	\begin{scriptsize}
		\begin{tabular*}{0.35\textwidth}{l|llllll}
			Lines		& HV lines  & MV lines  & LV lines\\ \hline 
			$C_j$ [MVA]			 & 130  & 14   &  0.12\\
			$l_j$ [km]		 & 62.7  	&	4.72		 &	0.17 \\
			$L_j$[1000 km]			   	 & 76 & 		523	&	 	1067.1 \\
			$U_j$ [kV]	&	110		&	30		&	0.4\\
			$R$ [$\Omega/km$]			&	0.1		&0.4	&	0.5\\
			$X$	[$\Omega/km$]	&	0.387		&0.3	&	0.08\\
			$Y_{cr}$ [$km/\Omega$]		&	2.983e-6		&2.9202e-6 &	2.669e-6\\
		\end{tabular*}
	\end{scriptsize}
	\caption{Parameters of conceptual power chain network where $C$ is the average transmission or transformer capacity,  $l$ the average power line length,  $U_L$ the lower voltage side, $U_r$ the ohmic voltage drop, $U_k$ the short-circuit voltage, $R$ the resistance, $X$ the reactance and $Y_{cr}$ the cross admittance. Parameters and transformer impedance calculation according to \cite{oeding2004}.}
	\label{tab:links}
\end{table}

\end{document}